\documentclass{aa}
\usepackage{graphicx}

\begin{document}
\title{Radio spectrum evolution and magnetic field 
in extreme GPS radio sources}
\subtitle{The case of RXJ1459+3337}
\author{
M. Orienti\inst{1,2,3} \and
D. Dallacasa\inst{2,3}
}
\offprints{M. Orienti}
\institute{
Instituto de Astrofisica de Canarias, 38200-La Laguna, Tenerife, Spain \and
Dipartimento di Astronomia, Universit\`a di Bologna, via Ranzani 1,
I-40127, Bologna, Italy \and 
Istituto di Radioastronomia -- INAF, via Gobetti 101, I-40129, Bologna,
Italy 
}
\date{Received \today; accepted ?}

\abstract
{}
{The knowledge of the properties of the youngest radio sources is very
  important in order to trace the earliest phase of the evolution of 
the radio emission. RXJ1459+3337, with its high turnover
  frequency ($\sim$ 25 GHz) provides a unique opportunity to study
  this class of extreme objects.}
{High-sensitivity multi-frequency VLA observations have been carried
  out to measure the flux-density with high accuracy, while
  multi-frequency VLBA observations were performed, aimed at determining the
  pc-scale structure. Archival ROSAT data have been used to infer the
  X-ray luminosity.}
{ The comparison between our new VLA data and those available in the
  literature shows a steady increment of the flux-density in the
  optically-thick part of the spectrum and a
  decrement of the turnover frequency. In the optically-thin regime,
  the source flux density has already started to decrease.
Such a variability can be
  explained in terms of an adiabatically-expanding homogeneous radio
  component.
The frequency range spanned by our VLBA observations, together with
  the resolution achieved, allows us to determine the source size and
  the turnover frequency, and then to derive 
the magnetic field
  directly from these observable quantities. The value obtained in this way
  is in good agreement with that computed assuming equipartition
  condition. A similar value is also obtained by comparing the
  radio and X-ray luminosities.}
{}
\titlerunning{Radio spectrum evolution in GPS sources}
\maketitle
\section{Introduction}

The radio emission of extragalactic sources is synchrotron
radiation produced by relativistic electrons with a power-law energy
distribution. The relativistic electrons produce a power-law radio
spectrum which
reaches its maximum in
correspondence to the turnover frequency $\nu_t$. \\
At frequencies below
$\nu_{t}$, the
spectrum turns over likely due to Synchrotron Self-Absorption (Snellen
et al. \cite{sn00}), although free-free absorption may also play a
role (Mutoh et al. \cite{mutoh02}, Kameno et al. \cite{kameno00}).\\
The evolutionary models proposed to interpret the various stages of the
life-cycle of radio sources relate the typical spectral peak
of very small and then compact region with age.
In the evolutionary scenario, the peak of the spectrum
progressively moves toward lower frequencies as the radio source
expands/grows.
In this framework, GHz-peaked spectrum (GPS) and compact
steep-spectrum (CSS) radio sources, the former with $\nu_{t} \sim 1\;
{\rm GHz}$ and the latter with $\nu_{t} \sim 100-500\; {\rm MHz} $, are
considered to represent early stages in the individual radio source
evolution.\\  
The timescale of the spectral evolution in extremely young radio
sources is very short, of the order of a few tens of years 
  (Dallacasa \cite{dd03}). 
This implies that the ideal targets to investigate 
how the radio spectrum evolves, and
that are the main mechanisms at work, must be sought among
sources with very high
turnover frequencies.
The existence of ``extreme'' GPS sources with
turnover frequency above 10 GHz, termed ``high frequency peakers''
(HFPs) by Dallacasa et al. (\cite{dd00}), is expected 
from all the radio source evolutionary models. 
Their detection is, however, quite difficult since they are
  short-lived objects evolving into GPS and then CSS sources, and
to observational
limitations preventing their selection. 
For a source to be recognized
as an extreme GPS object, the characteristic turnover frequency
must lie within
the frequency range sampled by large area surveys. 
The frequencies of
the surveys currently available are not high enough to allow an
efficient selection of high-frequency peaking objects, since there is
insufficient information above $\sim$ 5 GHz.\\
\begin{figure*}[!h]
\begin{center}
\includegraphics{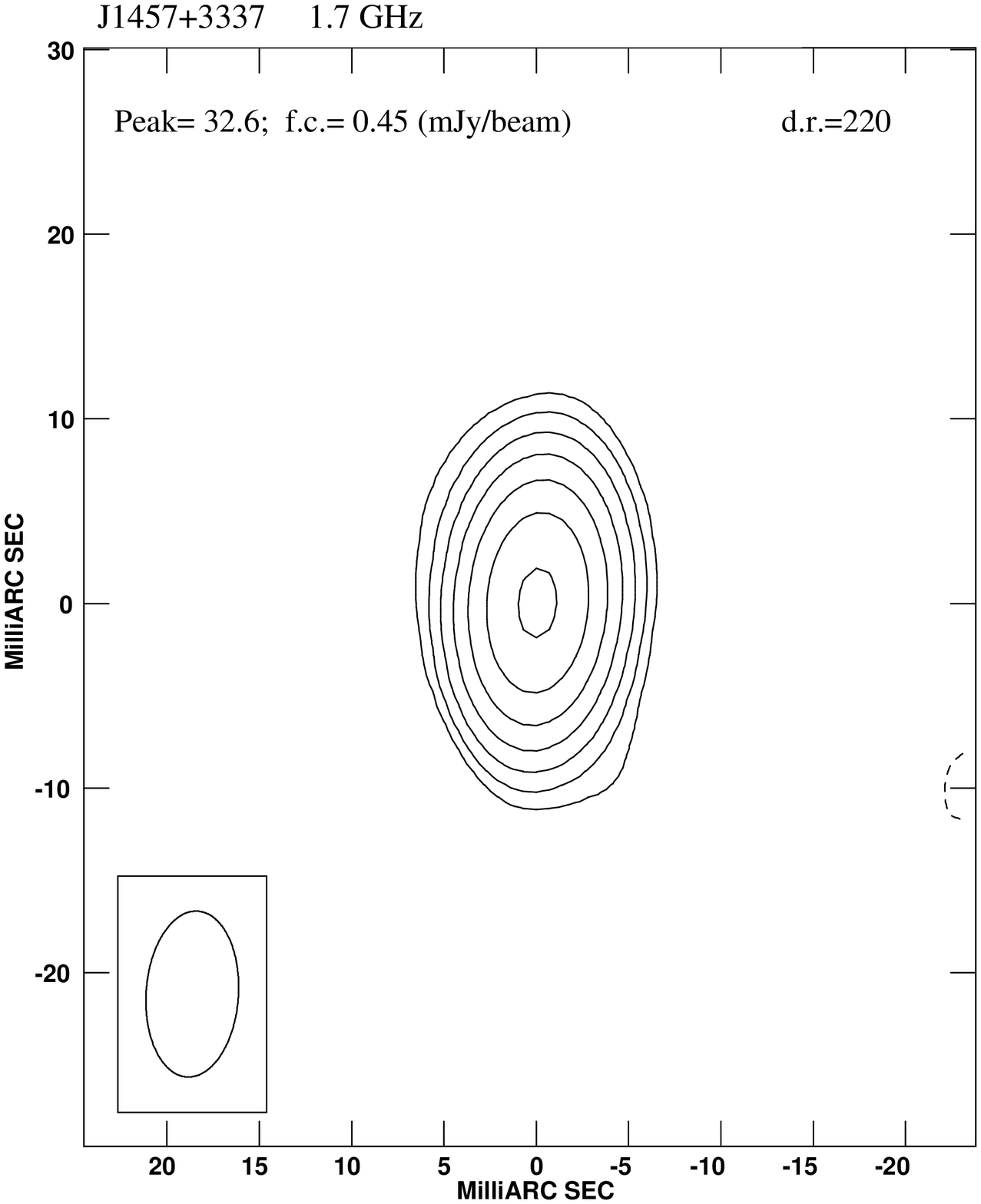}
\includegraphics{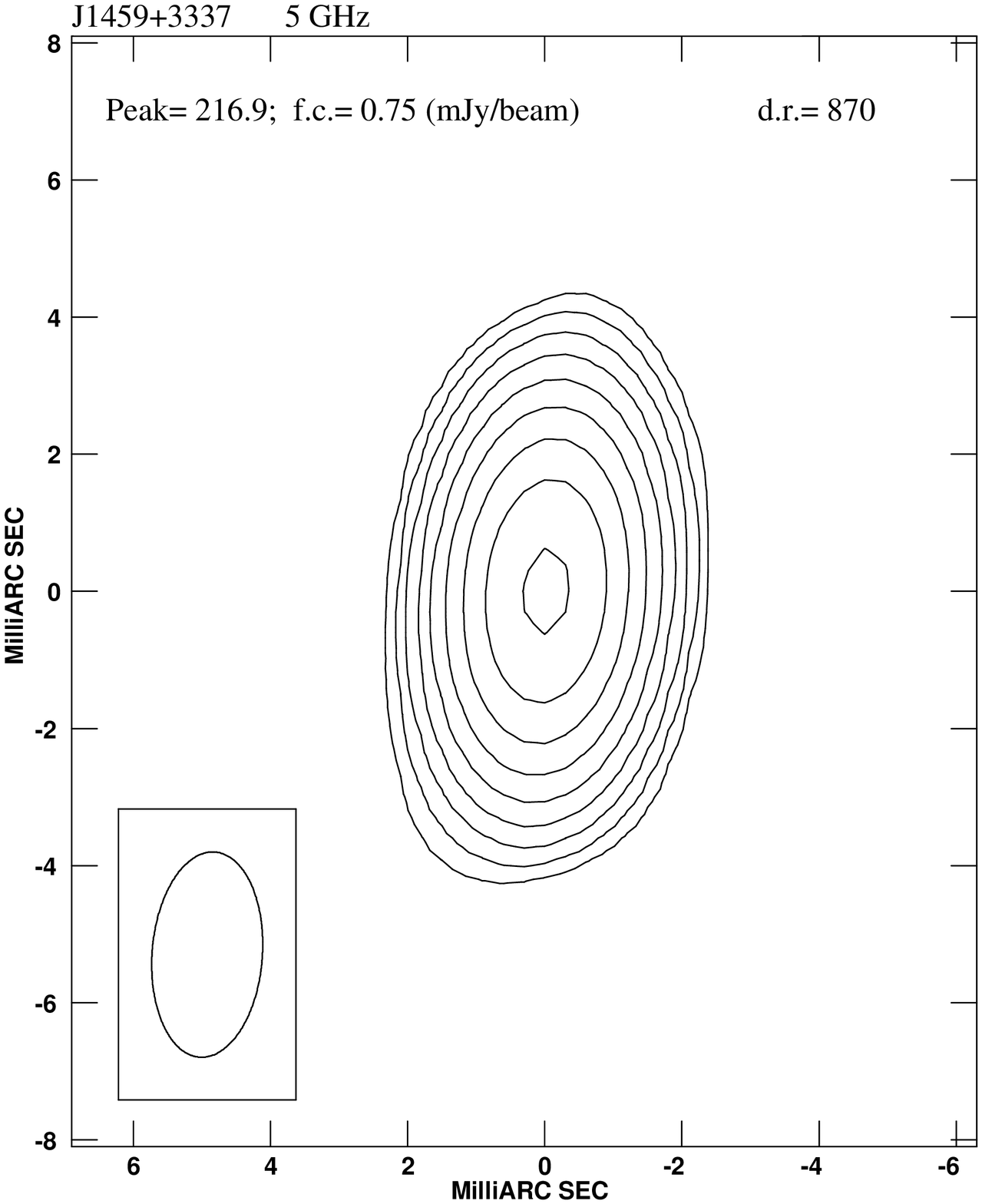}
\includegraphics{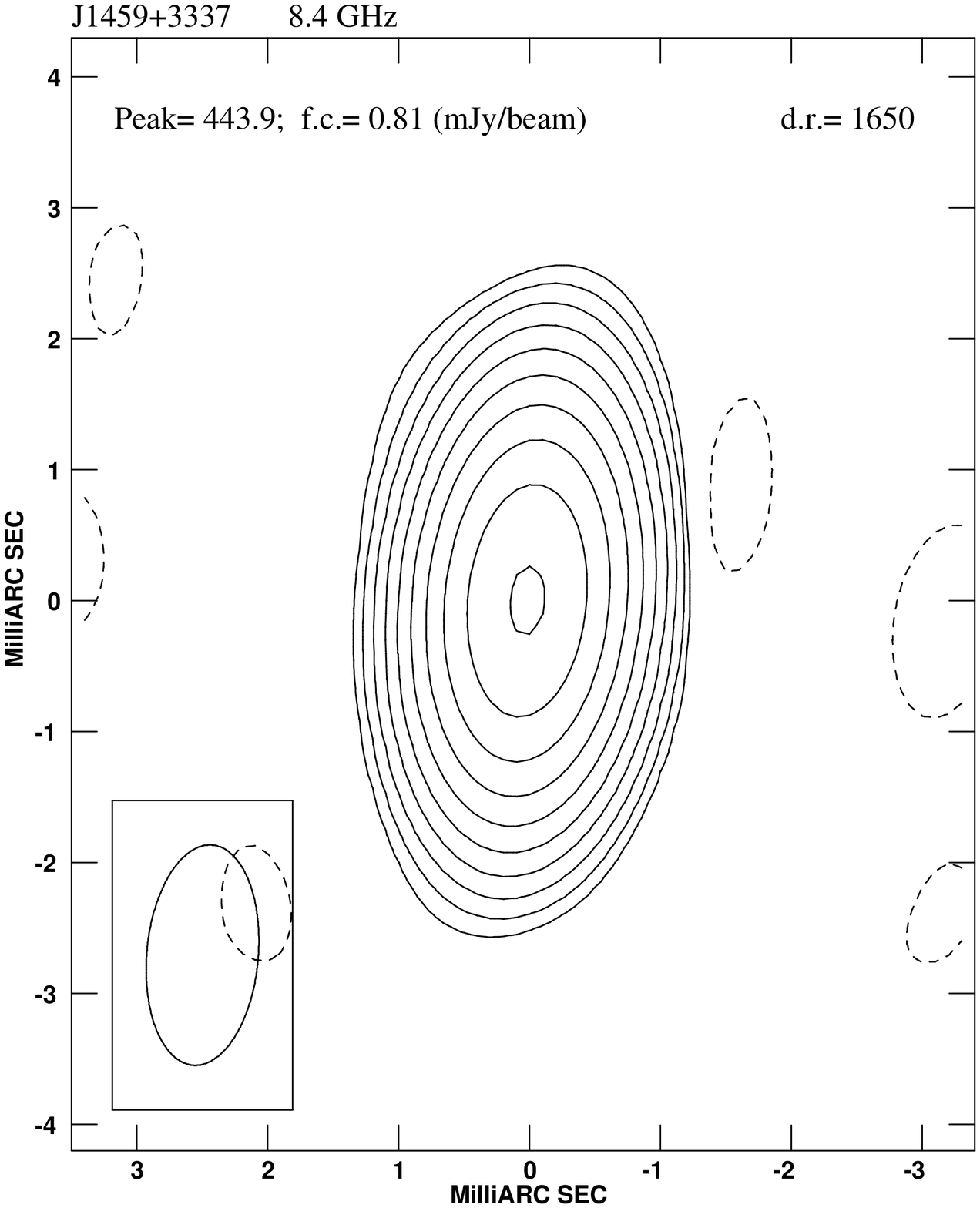}
\includegraphics{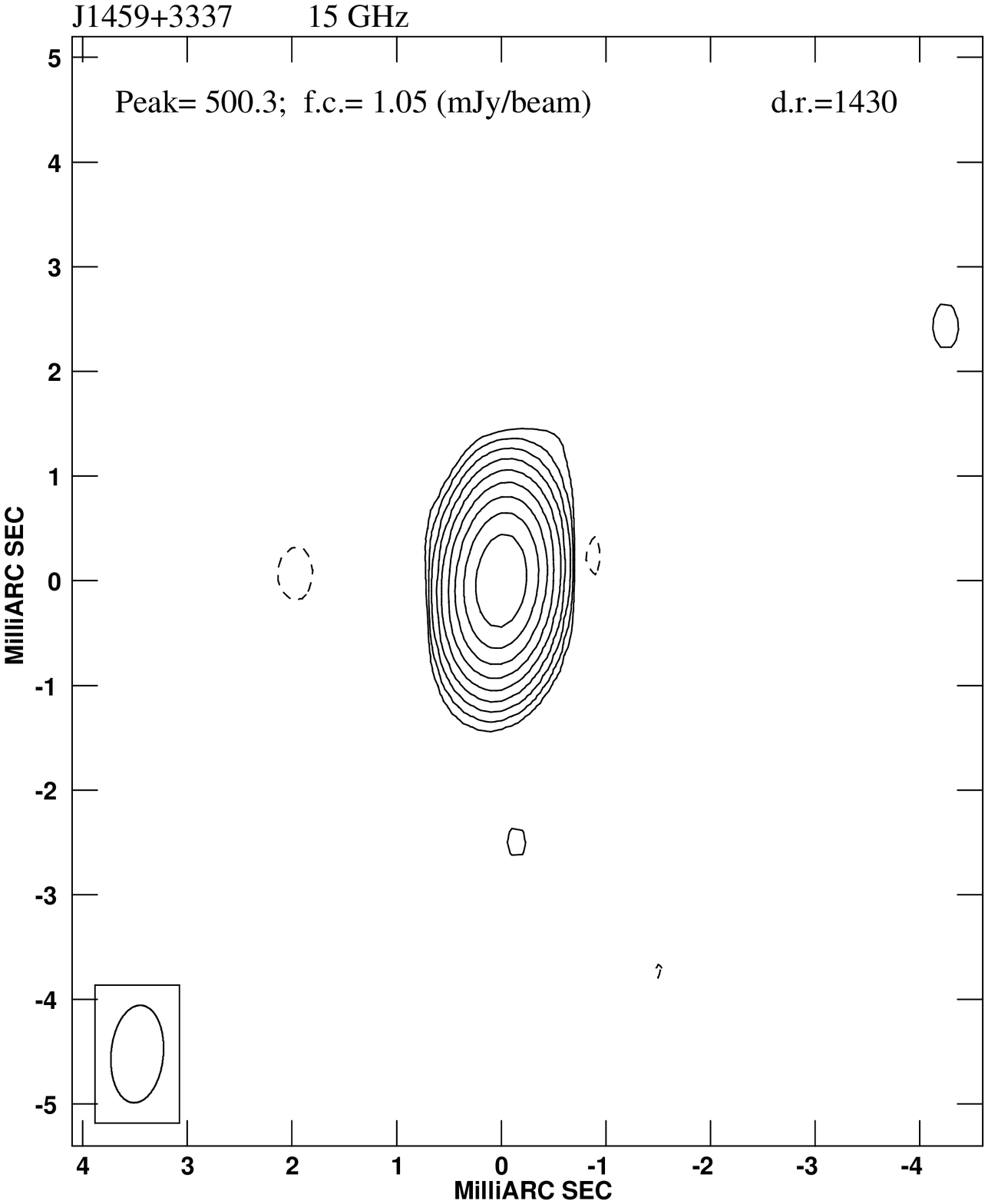}
\includegraphics{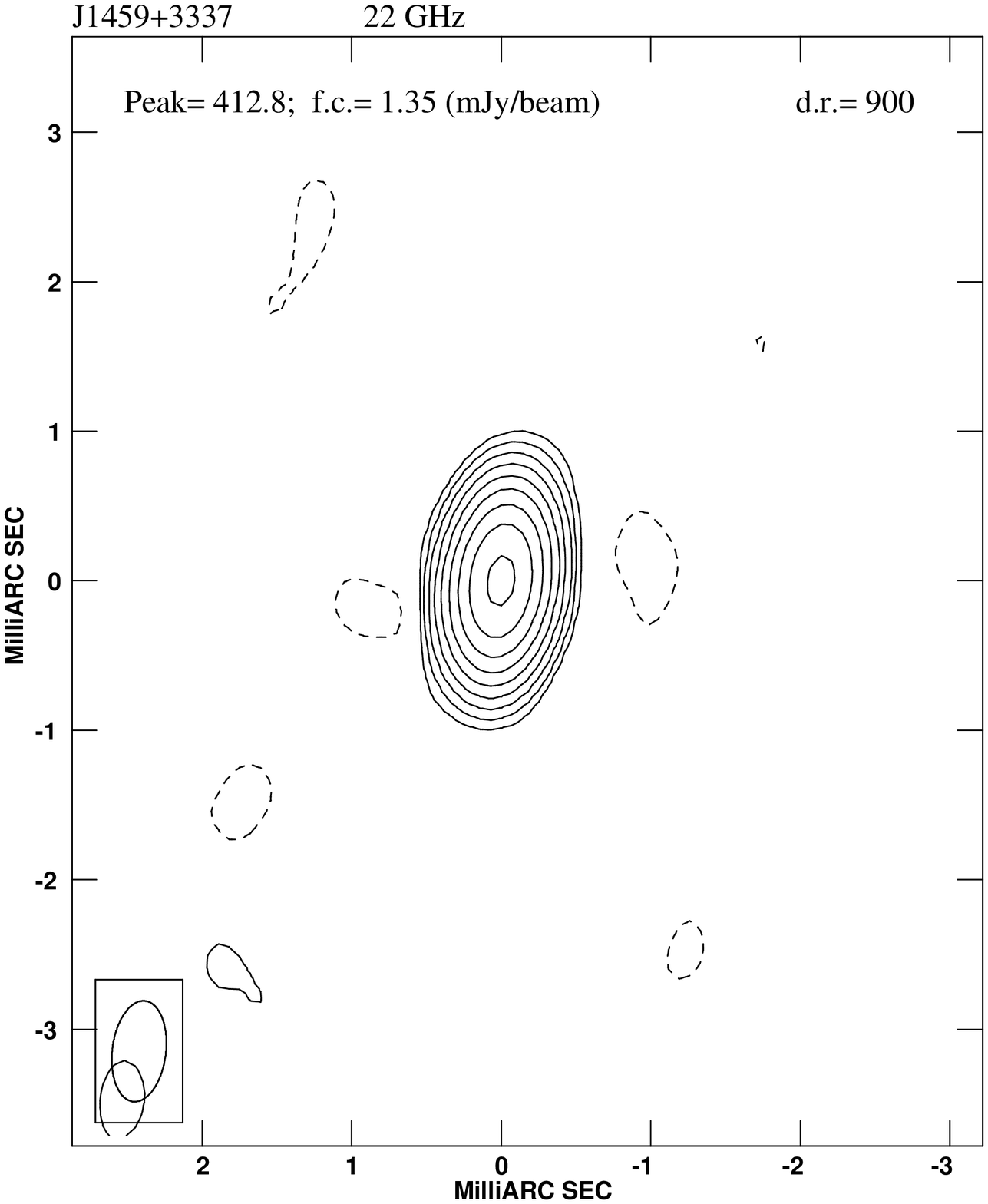}
\includegraphics{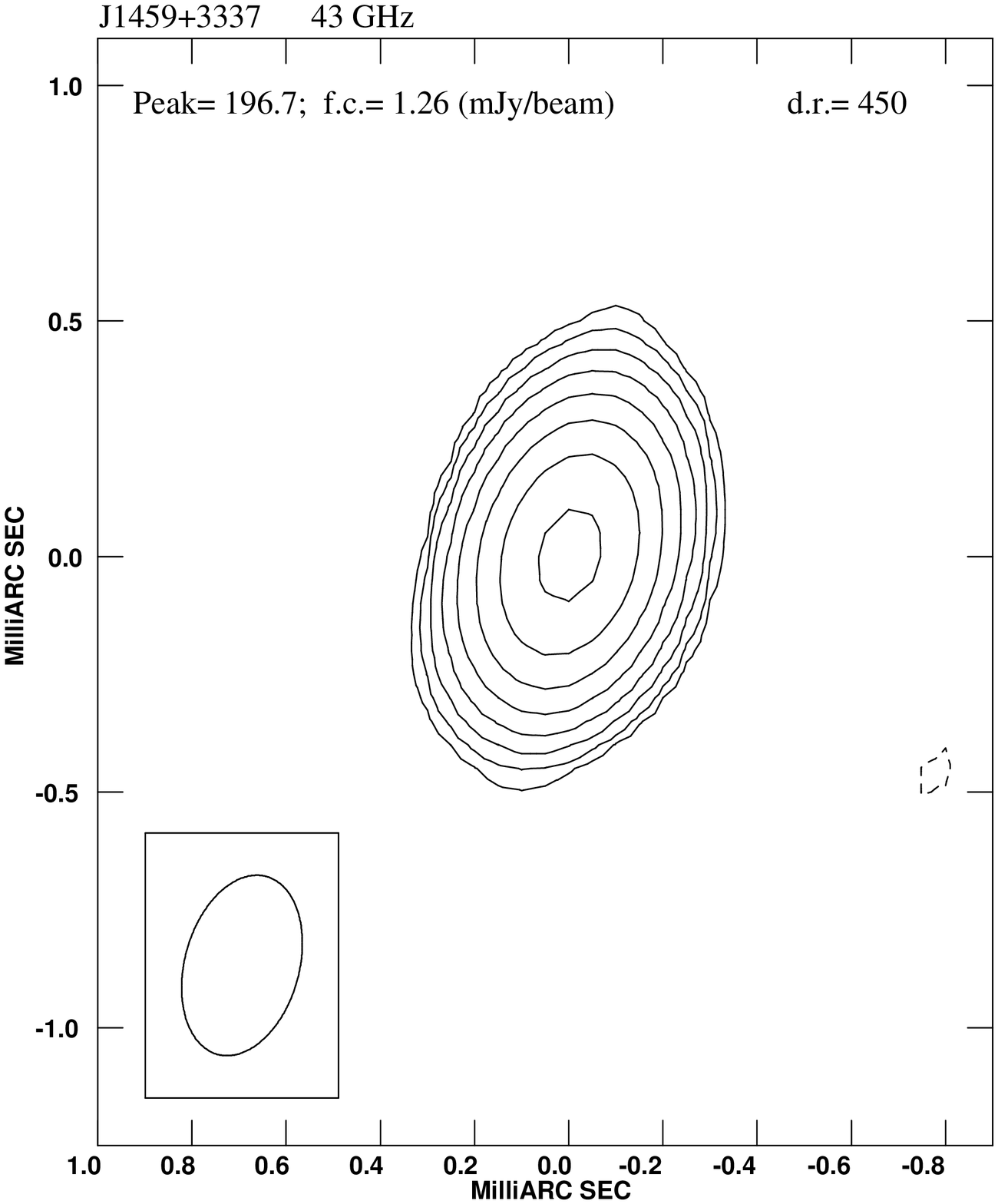}
\vspace{19cm}
\caption{VLBA images of the source J1459+3337 at 1.7, 5, 8.4, 15, 22 and
  43 GHz. On each image we give the following information on the plot
  itself: a) Peak flux density in mJy/beam; 
b) First contour intensity (mJy/beam), which is generally 
equal to 3 r.m.s. of the noise level; 
contour levels increase by a factor of 2; c) the dynamic range 
  (the ratio of the peak brightness and 1$\sigma$);
d) The restoring beam, plotted on the bottom left corner of each image. }
\label{immagini}
\end{center}
\end{figure*}
The radio source RXJ1459+3337 was one of the first objects to be
recognized as an ``extreme'' GPS with the
turnover frequency at about 30 GHz and a peak flux density of
  about 800 mJy as reported by Edge et al. (\cite{edge96}). 
This object is associated with a quasar at z = 0.6448. It was
detected by ROSAT (Brinkmann et al. \cite{brink00}) and shows
an X-ray luminosity $L_{X} \sim 2.8\times 10^{45} {\rm erg/s}$.\\ 
Multi-epoch VLA observations  
indicate strong variability in the optically-thick part of its
radio spectrum. In particular, the flux-density variability at 5 GHz
(i.e. in the optically-thick regime) seems
to steadily increase. 
Although flux-density variability is quite rare in GPS sources (O'Dea
\cite{odea98}), an increment of the flux density at frequencies below
$\nu_{t}$ may be a consequence of the evolution of a synchrotron
self-absorbed spectrum in an expanding component.\\
In this paper we present the results of 
multi-frequency VLA and VLBA observations conducted in 2003 and
  2005, respectively. In order to
  identify the mechanisms at the basis of the spectral evolution we
  combine the information on the physical conditions provided by 
  our new VLA data with that from archival data obtained in 1996
    and 1999. 
The frequency range sampled by
the new VLBA observations, together with their high-resolution, 
allows us to compute the magnetic field by means of
observable quantities only, such as peak frequency, peak flux
  density and the source size, as from the synchrotron theory
  (Kellermann \& Pauliny-Toth \cite{kpt81}; see also Section 4.2). 
The comparison between this value with the
field strength obtained assuming equipartition provides important
information on the physical conditions of the radio source.

Throughout this paper, we assume the following cosmology: $H_{0} = 71\; {\rm
  km\,s^{-1}\,Mpc^{-1}}$, $\Omega_{\rm M} = 0.27$ and $\Omega_{\lambda}
  = 0.73$, in a flat Universe.\\

\section{Observations and data reduction}

\begin{figure}
\begin{center}
\includegraphics{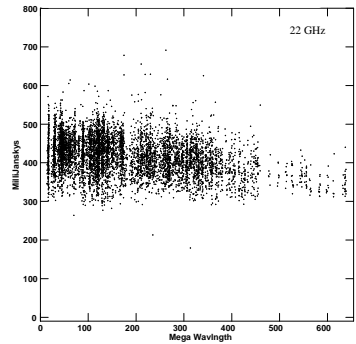}
\vspace{5cm}
\caption{Plot of amplitude vs projected baseline length at 22 GHz.}
\label{uvplot}
\end{center}
\end{figure}

Our target RXJ1459+3337 was observed with the VLA on September 12th 2003
at 8 independent frequencies (1.4, 1.7, 4.5, 4.9, 8.1, 8.4, 15.0 and
22.2 GHz) during the monitoring program of HFP candidates 
(Dallacasa et
al. \cite{dd00}; Tinti et al. \cite{st05}), 
the characteristics of the observations and the
data reduction are described in Orienti et al. (\cite{mo07}).
In VLA data the r.m.s noise level on the image plane is not relevant
(always well below 1 mJy),
if compared to the main
uncertainty coming from the amplitude calibration errors, which are
within (1$\sigma$) 3\% at 1.4, 1.7, 4.5, 5.0, 8.1 and 8.4 GHz, and 5\%
at 15 and 22 GHz, and which are predominant in the case of relatively
strong radio sources such as RXJ1459+3337.\\
VLBA observations of the radio source RXJ1459+3337 were carried
out on April 4th 2005 at 1.7, 5, 8.4, 15, 22 and 43 GHz, in
full polarization mode with a recording band-width of 32 MHz at 128
Mbps, for a total time of 10 hours. 
The correlation was performed at the VLBA correlator in
Socorro and the data reduction was carried out with the NRAO AIPS package.
After the application of system temperature and antenna gain, the
amplitudes were checked using the data on 4C39.25 (J0927+3902).
The error on the absolute flux density scale is generally within
3\%--10\%, being worse at the highest frequencies.
The same source 4C39.25 was used to generate the bandpass
correction at each frequency.
However, at 43.2 GHz 
a problem on the scan on the calibrator 
precluded a good bandpass calibration.\\
Images of RXJ1459+3337 at each frequency were produced after a number of
phase-only self-calibration iterations (Fig. \ref{immagini}). 
The source was found to
  be marginally
resolved at all frequencies (Fig. \ref{uvplot}).
\begin{table*}
\begin{center}
\begin{tabular}{ccccccccccc}
\hline
\hline
&&&&&&&&&&\\
 &Date&$S_{1.4}$&$S_{1.7}$&$S_{4.5}$&$S_{5.0}$&$S_{8.1}$&$S_{8.4}$&$S_{15}$&$S_{22}$&$S_{43}$\\
 & &mJy&mJy&mJy&mJy&mJy&mJy&mJy&mJy&mJy\\
\hline
&&&&&&&&&&\\
VLA&Sept 2003&25&31&194&221&410&423&521&356&\\
VLBA&Apr 2005&  &33&   &221&   &446&519&348&198\\
&&&&&&&&&&\\
\hline
\end{tabular} 
\vspace{0.5cm}
\end{center}   
\caption{The VLA and VLBA flux densities of RXJ1459+3337.}
\label{parameter}
\end{table*}  
The source flux densities at each frequency 
have been measured on both VLBA and VLA images 
in several ways, giving similar results, and the values 
are reported in Table \ref{parameter}. 
The VLBA flux-densities at each frequency are 
consistent within $< 10\%$ with those measured by the VLA, indicating the 
lack of low-surface brightness features on the pc-scale.\\ 

\section{Results}  

\begin{figure*}[th]
\begin{center}
\includegraphics{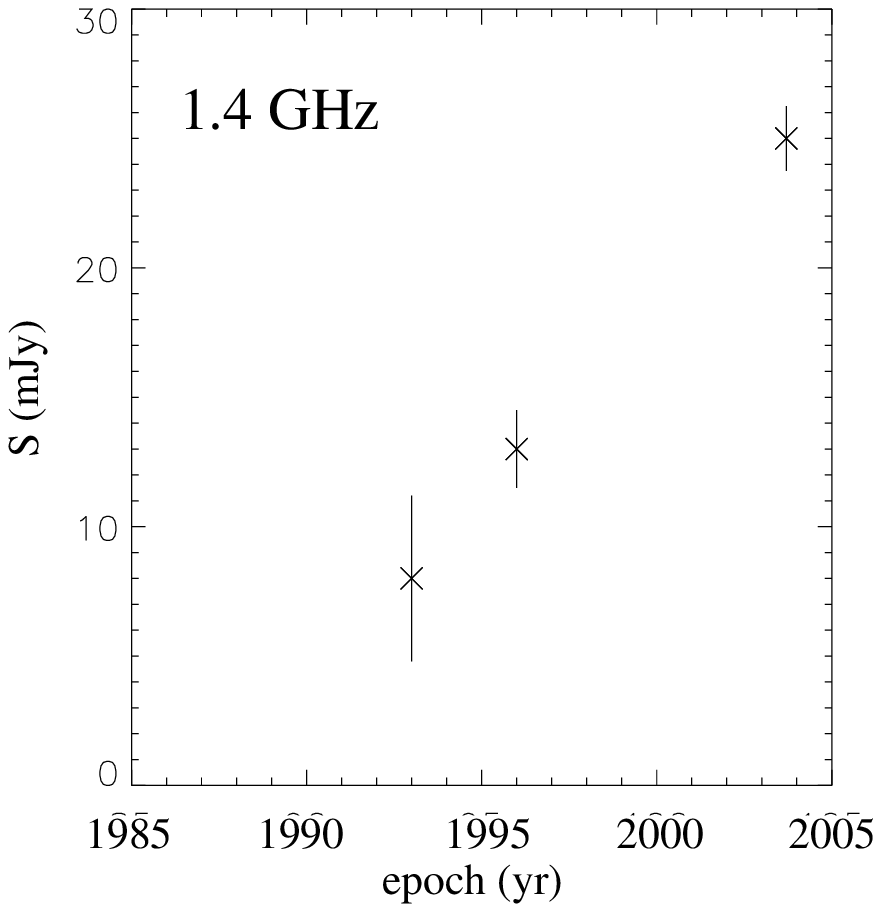}
\includegraphics{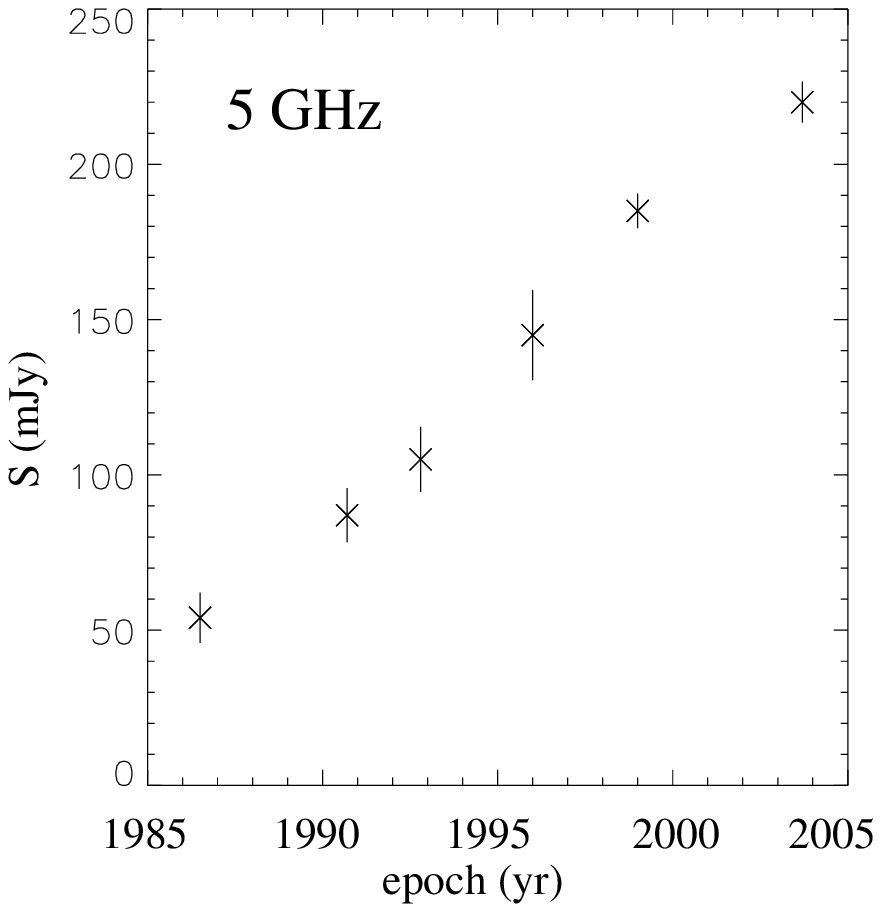}
\includegraphics{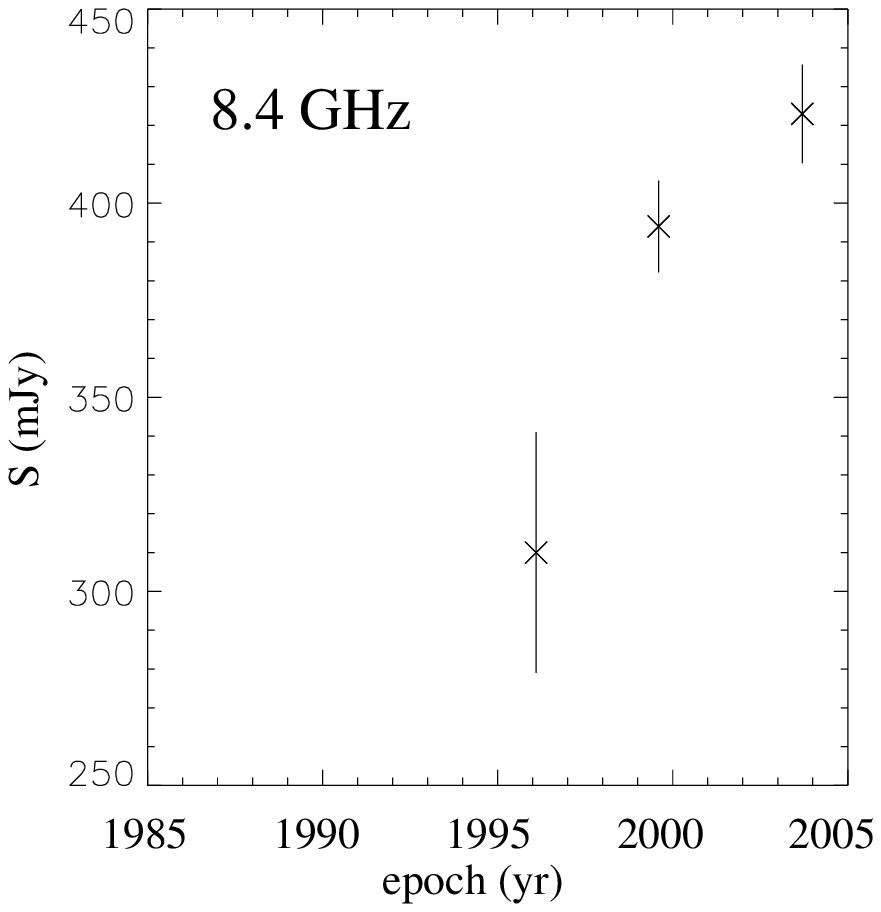}
\includegraphics{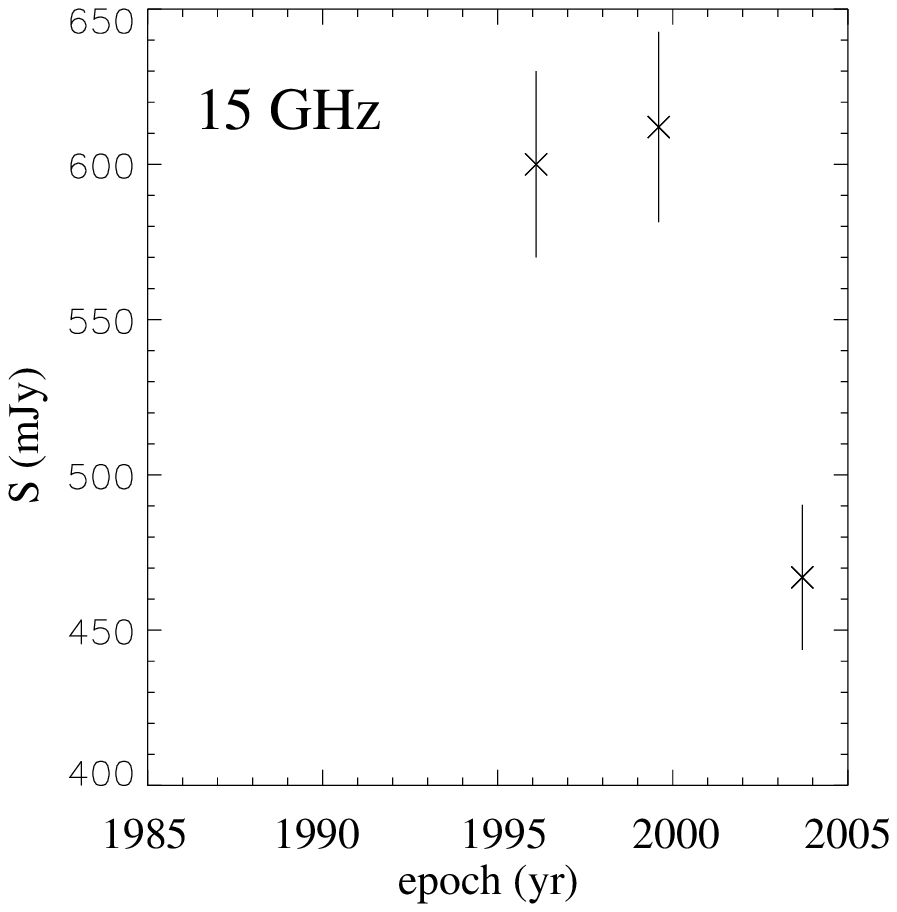}
\includegraphics{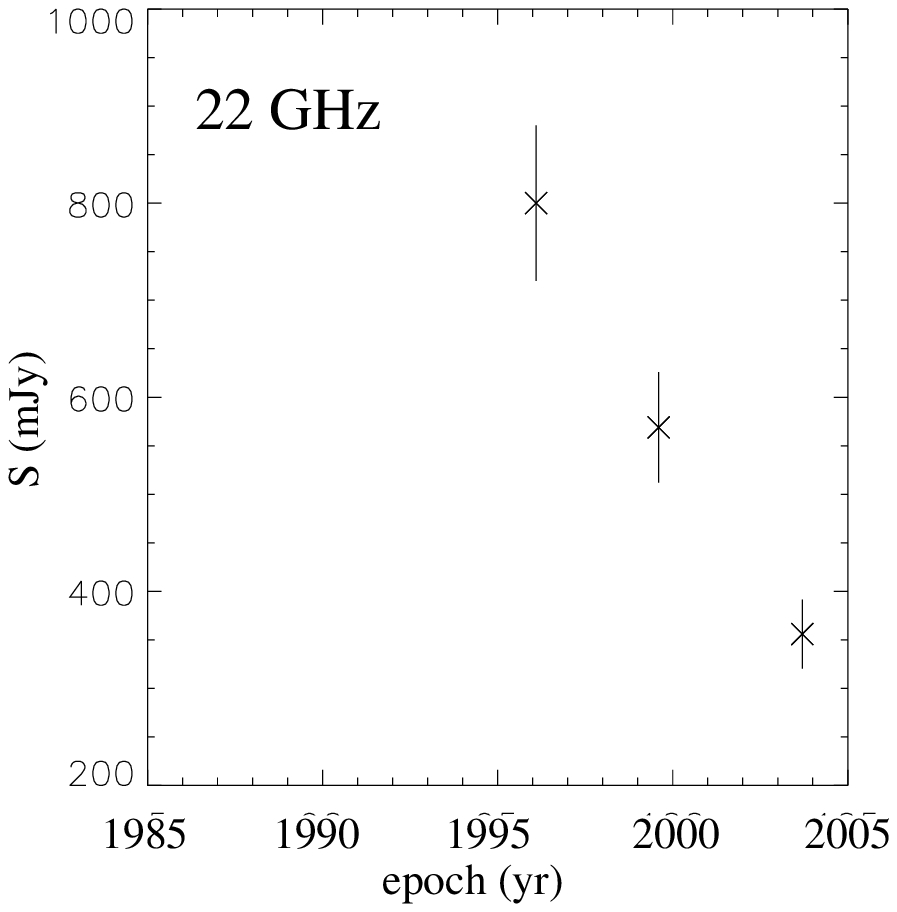}
\vspace{13cm}
\caption{Light-curves of
  RXJ1459+3337 at 1.4, 5, 8.4, 15 and 22 GHz. 
In addition to the data points presented in this paper we include 
  those concerning the epochs 1986 (87GB, Gregory et
  al. \cite{gregory96}), 
  1991 (Neumann et al. \cite{neumann94}), 1993 (Laurent-Muehleisen et
  al. (\cite{laurent97}), 1996 (Edge et al. \cite{edge96})
  and 1999 (Dallacasa \cite{dd03}).}
\label{var_vla}
\end{center}
\end{figure*}

GPS radio sources are usually characterized by the lack of any
significant flux-density variability and they can be considered as the
least variable class of extragalactic radio sources (O'Dea
\cite{odea98}), with an average variation within $\sim$5\% (Stanghellini et
al. \cite{cs05}). In the case of RXJ1459+3337, however,
a comparison between our new simultaneous multi-frequency VLA
observations  
with those available from the literature has pinpointed a
substantial variability in the optically thick part of the spectrum.\\
Fig. \ref{var_vla} shows the light-curves of RXJ1459+3337 at
  each frequency:
from these plots it seems that at 
1.4 and 5 GHz, well below the turnover frequency, the flux density 
has steadily increased, while 
in the optically-thin regime {(i.e. 22 GHz)}, the flux density has
decreased. Following the approach from Dallacasa
et al. (\cite{dd00}), we fit the simultaneous radio spectra with a
purely analytical function used to determine 
the peak flux density $S_{m}$ and the frequency $\nu_{m}$
at which it occurs:

\begin{displaymath}
{\rm Log}S = a - \sqrt{b^{2} + (c{\rm Log}(\nu ) - d)^{2}} 
\end{displaymath}

\begin{figure}
\begin{center}
\includegraphics{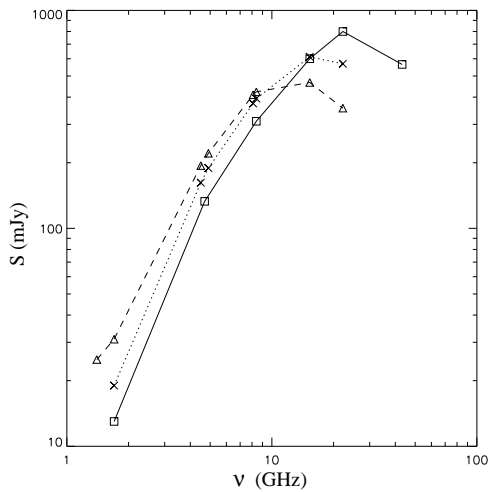}
\vspace{6.5cm}
\caption{The VLA radio spectrum of RXJ1459+3337 in 1996 
    ({\it squares}, Edge et al. \cite{edge96}), in 1999 ({\it
    crosses}, Dallacasa \cite{dd03}), and in 2003 ({\it triangles},
    these new data).}
\label{spectra}
\end{center}
\end{figure}

\noindent 
In this equation, the parameters $a,b,c$ and $d$ are 
purely numeric, and do not provide any direct physical information.
Based on the results of the fit, we found that
the spectral peak has moved to lower
frequencies, from $\sim$24$\pm$1 GHz in 1996 to 17$\pm$2 GHz in
  1999 and to 12.5$\pm$0.5 in
2003 (Fig. \ref{spectra}). The errors on these quantities were
  calculated following the error propagation theory.\\
The decrement of the spectral peak together with the steady
increment of the flux-density in the optically-thick part of spectrum
suggest that the source is adiabatically expanding.
Despite the high resolution achieved by VLBA observations, the
radio source RXJ1459+3337 is only marginally resolved 
even at 43 GHz.
Since the inferred source size is substantially smaller than the
  VLBA beam
size at the highest frequency, our measurements are not sensitive to
small changes on the angular size and thus cannot confirm the
  source expansion suggested by the spectrum evolution.\\

\section{Discussion}

The strong variability in the optically thick part of the spectrum is
not common among sources peaking below 5 GHz (O'Dea \cite{odea98}).
However, the timescale of evolution of {\it extreme} GPS sources
appears to be
quite short, of the order of a few tens of years.
In the context of
the youth scenario, such a result is not unexpected since
very young radio sources should evolve quite rapidly (Dallacasa
  \cite{dd03}).\\
In the following discussion we investigate the physical conditions of
the radio emission and which mechanisms are involved in the evolution of
the radio spectrum.\\

\subsection{The evolution of the radio spectrum}

As noted above, 
the radio emission in extragalactic radio sources is due to
synchrotron radiation from relativistic particles with a power-law
energy distribution. However, in very small objects the observed radio spectrum
significantly departs from the power-law shape, turning over at
frequencies below the peak. Such a deviation is mostly explained in
terms of synchrotron self-absorption.\\
The mechanism which plays the major role in the evolution of the
optically-thick part of the spectrum is the adiabatic expansion.
In this regime we have that 

\begin{equation}
S(\nu) \propto H^{-\frac{1}{2}} \nu^{\frac{5}{2}} \theta^{2}
\label{ssa}
\end{equation}

\noindent (Kellermann \& Pauliny-Toth \cite{kpt81})
where $S$ is the flux density at a given frequency $\nu$ in the
optically-thick part of the spectrum, $H$ is the magnetic field 
and $\theta$ is the angular size of the emitting region.
As the radio source adiabatically expands, the opacity decreases, the
turnover frequency moves to lower frequencies and the flux
density increases at frequencies below the turnover frequency.\\
Is it possible to relate the flux-density variability observed in
RXJ1459+3337 to adiabatic expansions? \\   
First of all, we assume that the radio emission is due to a
homogeneous component which is adiabatically expanding at a constant
rate:\\

\begin{displaymath}
\theta = \theta_{0} \left( \frac{t_{0} + \Delta t}{t_{0}} \right)
\end{displaymath}

\noindent where $\theta_{0}$ is the source angular size at the time
$t_{0}$ and $\theta$ at the time $t_{0} + \Delta t$.\\
We also assume that the magnetic field is frozen in the plasma:\\

\begin{displaymath}
H = H_{0} \left( \frac{t_{0}}{t_{0}+\Delta t} \right)^{2}
\end{displaymath}

\noindent where $H_{0}$ is the magnetic field at the time $t_{0}$ and
$H$ the magnetic field at the time $t_{0} + \Delta t$.
\noindent From Eq. \ref{ssa} we find that:\\

\begin{equation}
\frac{S}{S_{0}} = \left( \frac{t_{0} + \Delta t}{t_{0}} \right)^{3}
\label{ratio_s}
\end{equation}

\noindent where $S_{0}$ and $S$ are the flux densities measured at the same
frequency at the time $t_{0}$ and $t_{0} + \Delta t$ respectively.\\
In the case of the 5 GHz light-curve, where enough data points are
available, we estimate the time of onset of radio emission by
fitting 
the flux density measured at the different epochs
with the function\\

\begin{displaymath}
y = b + (x/a)
\end{displaymath}

\noindent where $y = (S/S_{0})^{1/3}$, $x = t - t_{0}$, $a$ the age
the radio emission had at epoch $t_{0}$ and $b$ a free parameter
  which is almost equal to unity in our fits. As $S_{0}$
we consider the flux density measured at the oldest epoch
available ($t_0$; i.e. 1986). From the best fit to the data we obtain
$a\sim 30\pm 5$ years, i.e. the birth of the radio emission should
have occurred in 1956$\pm$5.
In the case of the 1.4 GHz light-curve, the availability of only 3
data points makes the determination of the age less accurate with $a
\sim 27\pm 15$ years, i.e. the radio emission should have started in
1966$\pm$15, in agreement with the 5 GHz determination.   
We do not perform fits on the light-curves at higher
frequencies since they are affected by contamination due to the 
flattening occurring near the spectral peak.

As previously mentioned, adiabatic expansion also causes the spectral peak
to shift at lower frequencies. We know that the turnover frequency
is\\

\begin{equation}
\nu_{t} \propto E^{2} H
\label{turnover}
\end{equation}
  
\noindent where the energy of the relativistic particles is
  defined as $E \propto t^{-1}$
(Pacholczyk \cite{pacho70}). This implies that

\begin{equation}
\frac{\nu_{0}}{\nu} = \left( \frac{t_{0} + \Delta t}{t_{0}}
\right)^{4}
\label{ratio_nu}
\end{equation}

\noindent where $\nu_{0}$ and $\nu$ are the turnover frequency at the time
$t_{0}$ and $t_{0} + \Delta t$ respectively.\\
We estimate the epoch of the origin of radio emission by fitting the
frequency peak measured at the different epochs with the same function used
to fit the multi-epoch flux density. In this case, y=($\nu_{0}$/$\nu$)$^{1/4}$,
$x= t - t_{0}$ 
and $a$ the age the radio emission had at the epoch $t_{0}$. $\nu_0$ is
the peak frequency at the oldest epoch available ($t_{0}$, in this case 1996).
From the best fit we find $a \sim 47\pm10$ years, implying that the radio
emission originated at $\sim1949\pm10$, still in agreement 
with the values
obtained from the flux density increment, although with higher
uncertainties, since the fit has been performed on 3 data points only.
The good agreement on the determination of the source age
may indicate that the radio source is evolving
with the physical conditions predicted 
by simple self-similar growth models, as previously assumed.\\

\subsection{The magnetic field}

Direct measurements of the magnetic field are very difficult to carry
out. The magnetic field $H$ can be determined
from observable quantities, such as the turnover frequency $\nu_{t}$ in GHz,
the peak flux density $S_{\rm max}$ in Jy and the source angular sizes
$\theta_{\rm maj}$ and $\theta_{\rm min}$
in mas as 
directly inferred from the observations.
In this case we have\\

\begin{equation}  
H \sim 10^{-4} \theta_{\rm maj}^{2} \theta_{\rm min}^{2} \nu_{t}^{5}
S_{\rm max}^{-2} (1+z)^{-1}    
\label{dir_b}
\end{equation}

\noindent (Kellermann \& Pauliny-Toth \cite{kpt81}) where $H$ is in Gauss.
From the fit to our new VLA data (see Section 2), 
we obtained $\nu_{t}$
$\sim$ 12.5$\pm$0.5 GHz and $S_{\rm max}$ $\sim$550$\pm$15 mJy. 
In addition, by using the task JMFIT we measured a source angular size of
  0.36$\times$0.14 mas ($\theta = 1.8 \times \theta_{\rm fit}$ where
  $\theta_{\rm fit}$ is the deconvolved full width at
    half-maximum of a Gaussian brightness
  distribution as used in the fitting procedure) on our best VLBA
  image, namely that at 22 GHz, since the longest spacings at 43 GHz are poorly
sampled and with a rather high noise. 
The resulting resolution at 43 GHz is therefore no better than 
that at 22 GHz.
With these parameters we obtain $H \sim 0.16\pm0.03$ G.\\
Another way to constrain the magnetic field is to assume that the radio
source is in equipartition conditions.  
From the minimum energy conditions (Pacholczyk \cite{pacho70}), 
the equipartition magnetic field is\\

\begin{equation}
H = \sqrt{ \frac{24}{7} \pi u_{\min}}
\label{h_equi}
\end{equation}

\noindent where the minimum energy density $u_{\rm min}$ is\\

\begin{displaymath}
u_{\rm min} = 7 \cdot 10^{-24} \cdot \left( \frac{L}{V}\right)^{4/7} \cdot (1+k)^{4/7}
\end{displaymath}

\noindent and $L$ is the luminosity calculated at 22 GHz, i.e.
in the
optically-thin part of the spectrum, and $V$ the volume of the radio
source in pc$^{3}$.
For the calculation of V, 
we assume that the radio emission has an ellipsoidal geometry and an
average optically thin spectral index of 0.7 (S $\propto \nu^{-
  \alpha}$). 
Furthermore, proton and
electron energies are assumed to be equal (k=1), with a filling factor
of
unity (i.e. the source volume is fully and homogeneously filled by
relativistic plasma). From Eq. \ref{h_equi},
we calculate an equipartition magnetic field $H_{\rm eq}$
$\sim$ 0.16 G, in very good agreement with the value directly derived
from the observed spectral quantities.\\
Since RXJ1459+3337 is also an X-ray source, we can try to constrain
the magnetic field by comparing the radio and X-ray luminosities, if we
assume that all the X-ray emission is due to Comptonization of the
electrons responsible for the synchrotron emission in the radio
band.
In this case we have:\\

\begin{equation}
\frac{L_{\rm SSC}}{L_{\rm Syn}} \sim \frac{8 \pi w_{f}}{H^{2}}
\label{ratio_l} 
\end{equation}

\noindent (e.g. Singal \cite{singal86}) 
where $L_{\rm Syn}$ and $L_{\rm SSC}$ are the synchrotron and the
synchrotron self-Compton luminosity respectively and $w_{f}$ is the
radiation field, defined as\\

\begin{displaymath}
w_{f} = \frac{3 L_{\rm Syn}}{Ac}
\end{displaymath}

\noindent with $A$ 
the surface area of the synchrotron emission,
and $c$ the
speed of light.
The ROSAT X-ray flux density of RXJ1459+3337 is 1.56$\times$10$^{-12}$
erg s$^{-1}$ cm$^{-2}$ (Brinkmann et al. \cite{brink00}), 
which corresponds to an X-ray
luminosity $L_{\rm SSC}$ $\sim$ 2.8$\times$10$^{45}$ erg/s.
From the radio flux density we derive a synchrotron luminosity $L_{\rm
Syn}$ $\sim$ 1.2$\times$10$^{45}$ erg/s.\\
If in Eq. \ref{ratio_l} we introduce these luminosities and the source
emitting area $A \sim 3.9\, {\rm pc^{2}}$ as derived 
from VLBA images assuming an
ellipsoidal geometry (see above), we obtain a magnetic
field H $\sim$ 0.19 G, still in good agreement with the values derived
by the previous methods.\\
These results strongly suggest that the radio emitting plasma in such an
extreme object is consistent with the equipartition conditions.\\

\section{Conclusions} 

We have presented new multi-frequency 
VLBA and VLA observations of the radio source
RXJ1459+3337. Given the position of the spectral peak at
very high frequency,  
this object is classified as an {\it extreme}
GPS radio source, also known as HFP.\\
By comparing our VLA observations with data from the literature spanning
about 17 years, we find that the flux-density in the
optically-thick part of the spectrum has been continuously increasing,
while the turnover frequency has been moving toward lower
frequencies.
Such a regular variability, although not common in GPS sources peaking below 5
GHz, can be described in the case the radio emission
originates from an adiabatically-expanding homogeneous component.\\
If in this context we compare either the flux-densities at the same frequency
measured at different times or the turnover frequencies, 
we constrain the age of the radio emission, which is about
$\sim$50$\pm$10 years. \\
The resolution achieved by the VLBA observations, together with the
frequency range sampled, allowed us to ``observe'' the
turnover frequency and to {\it directly} measure parameters, such
as the peak flux density and angular size of the source with great
accuracy. 
In this
way we estimate the magnetic field to be 0.16 $\pm$ 0.03 G 
by means of observable
quantities only. 
The consistency of this value with the field derived for a minimum
energy condition 
strongly supports the idea that
such extreme objects, as RXJ1459+3337 may already be 
in equipartition.
The availability of ROSAT X-ray observations allowed us to infer the
magnetic field by comparing radio and X-ray luminosities. Even in this
case, the value inferred for the magnetic field agrees with the
previous results.\\

\begin{acknowledgements}
We thank the anonymous referee for carefully reading the manuscript
and valuable suggestions.
The VLA and VLBA are operated by the US National Radio Astronomy Observatory
which is a facility of the National Science Foundation operated under
a cooperative agreement by Associated University, Inc. 
This work has
made use of the NASA/IPAC Extragalactic Database NED which is operated
by the JPL, California Institute of Technology, under contract with
the National Aeronautics and Space Administration.  
\end{acknowledgements}

\end{document}